# Discovery of a Highly Anisotropic Type-II Ferromagnetic Weyl State Exhibiting a 3D Quantum Hall Effect

Yingdong Guan[1+], Abhinava Chatterjee[1+], Trace Bivens[2], Seng Huat Lee[3], Asuka Honma[4], Hirofumi Oka[5], Jorge D Vega Bazantes[6], Ruiqi Zhang[6], David Graf[7], Jianwei Sun[6], Seigo Souma[5], Takafumi Sato[4,5], Yong P. Chen[5,8,9], Yuanxi Wang[2], Chaoxing Liu[1*], and Zhiqiang Mao[1,5*]

[1] *Department of Physics, Pennsylvania State University, University Park, PA 16802, USA*
[2] *Department of Physics, University of North Texas, Denton, Texas 76201, USA*
[3] *2D Crystal Consortium, Materials Research Institute, The Pennsylvania State University, University Park, Pennsylvania 16802, USA*
[4] *Department of Physics, Graduate School of Science, Tohoku University, Sendai 980-8578, Japan*
[5] *Advanced Institute for Materials Research (WPI-AIMR), Tohoku University, Sendai, Japan*
[6] *Department of Physics and Engineering Physics, Tulane University, New Orleans, LA 70118, USA*
[7] *National High Magnetic Field Lab, Tallahassee, Florida 32310, USA*
[8] *Department of Physics and Astronomy and Elmore Family School of Electrical and Computer Engineering and Birck Nanotechnology Center and Purdue Quantum Science and Engineering Institute, Purdue University, West Lafayette, IN, USA*
[9] *Institute for Physics and Astronomy and Villum Center for Hybrid Quantum Materials and Devices, Aarhus University, Aarhus-C, 8000 Denmark*

## Abstract

**Topological semimetals, particularly Weyl semimetals (WSMs), are crucial platforms for exploring emergent quantum phenomena due to their unique electronic structures and potential to transition into various topological phases. In this study, we report the discovery of a ferromagnetic (FM) type-II WSM in $Mn(Bi_{1-x}Sb_x)_4Te_7$, which exhibits a remarkable three-dimensional (3D) quantum Hall effect (QHE). By precisely tuning the chemical potential through Sb doping, we obtained samples with the Fermi level near the charge neutrality point for $x \approx 0.27$. This was confirmed by spectroscopy measurements (ARPES and STS), and these samples showed strong quantum oscillations along with a key transport signature of a Weyl state - chiral anomaly, and Fermi surface reconstruction driven by FM ordering. Our theoretical analysis indicates that this Weyl state evolves from a parent nodal ring state, where higher-order k-terms split the nodal line into type-II Weyl nodes. The Weyl state exhibits significant anisotropy, characterized by a pronounced reduction in Fermi velocity along the**

**$k_z$-axis, likely accounting for the observed 3D QHE. These results not only highlight the exceptional tunability of the Mn(Bi$_{1-x}$Sb$_x$)$_4$Te$_7$ system, where precise control of the chemical potential and magnetic properties opens access to novel quantum phases, but also advance the understanding of FM WSMs.**

+Y.D.G. and A.C. equally contributed to this work

*email: cxl56@psu.edu; zim1@psu.edu

## I. INTRODUCTION

Topological insulators (TIs) and semimetals have been key subjects of interest in condensed matter physics due to their potential to give rise to emergent topological quantum phenomena with substantial technological significance [1-3]. Weyl semimetals (WSMs), a subset of topological semimetals, are distinguished by the presence of Weyl fermions as their low-energy excitations [4-13]. There are two established categories of Weyl semimetals: Type-I WSMs, characterized by untilted Weyl cones, were first discovered in the TaAs class of materials [14-16], while type-II WSMs, which feature strongly tilted Weyl cones that violate Lorentz invariance, were initially found in (W/Mo)Te$_2$ and later in a few other materials [17-23]. Recently, a variety of new classes of WSMs have also been proposed, such as higher-order WSMs [24], higher-order double WSMs [25], multifold WSMs [26-29], *etc*. One particularly intriguing feature of WSMs is their potential to transition into other topological phases under certain conditions. For instance, a three-dimensional ferromagnetic (FM) WSM is predicted to transform into a quantum anomalous Hall insulator (QAHI) when reduced to two dimensions [12,30]. Recent work by Gooth *et al.* demonstrated that axion insulators can be realized through charge-density-wave (CDW) modulated WSMs [31]. Thus, the discovery of new types of WSMs is highly desirable, as they could provide platforms for exploring exotic topological quantum states.

In this work, we report the observation of a novel ferromagnetic (FM) type II WSM in van der Waals (vd) layered and Sb-doped MnBi$_4$Te$_7$, which exhibits a remarkable three-dimensional quantum Hall effect (QHE) due to its highly anisotropic band structure and exceptionally low Fermi velocity along the $k_z$ direction. Pristine MnBi$_4$Te$_7$ belongs to the (MnBi$_2$Te$_4$)(Bi$_2$Te$_3$)$_n$ ($n$ = 0, 1, 2, 3, ...) material family, which has recently attracted significant interest. This is due to the combination of magnetism and non-trivial band topology in the system, offering access to a variety of topological quantum states. The $n$ = 0 member, MnBi$_2$Te$_4$, has been the most extensively studied. Shortly after being identified as an intrinsic antiferromagnetic (AFM) TI [32-34], numerous other exotic phenomena were discovered in its 2D thin layers, including the QAHI [35], axion insulator [36], high-Chern number insulator [37,38], layered Hall effect [39], and quantum metric-induced nonlinear Hall effect [40,41]. Moreover, when the magnetic phase of MnBi$_2$Te$_4$ is switched from AFM to FM, its bulk TI state is predicted to evolve into an ideal WSM state with a single pair of (type-II) Weyl nodes [32,34], a finding confirmed in prior experiments [42,43]. While MnBi$_2$Te$_4$ does not naturally exhibit a spontaneous FM phase, it can be induced via an external magnetic field. By carefully tuning the chemical potential through Sb doping at the Bi-site, Lee *et al.* detected transport signatures (chiral anomaly and a large anomalous Hall effect) of the predicted ideal FM Weyl state in Mn(Bi$_{1-x}$Sb$_x$)$_2$Te$_4$ (with $x \approx 0.26$) [42], where the chemical potential is near the charge neutrality point. The Fermi surface evolution revealed in recent quantum oscillation studies provides further evidence for the type-II nature of such a field-induced WSM [43]. In contrast to MnBi$_2$Te$_4$, which consists of Te-Bi-Te-Mn-Te-Bi-Te septuple layers (SL), MnBi$_4$Te$_7$ features a distinct heterostructure, formed from alternating stacks of MnBi$_2$Te$_4$ SLs and Bi$_2$Te$_3$ quintuple layers (QL). Like MnBi$_2$Te$_4$, MnBi$_4$Te$_7$ is also an intrinsic AFM TI [44] and is predicted to host a range of exotic topological states, including a high Chern number quantum spin Hall insulator and a QAHI in its 2D thin layers [45,46], a Möbius insulator in a canted AFM configuration [47], and type I or type II WSM states in the FM phase [48,49].

Another distinct characteristic of MnBi$_4$Te$_7$ is the increased spacing between its magnetic layers, which intensifies the competition between interlayer AFM and FM coupling [50-52]. The energy difference between the FM and AFM phases is minimal, with $E_{\text{AFM}} - E_{\text{FM}}$ = -0.23 meV [53], indicating a delicate balance between these magnetic states. This competition of magnetic interactions manifests in an AFM state below 13 K, followed by FM behavior below 5 K [49], enabling tunable magnetism in the compound through Sb doping [48,54,55]. Previous studies have shown that Sb-doped $MnBi_4Te_7$ can exhibit either a single FM transition upon lowering the temperature or an initial AFM transition followed by a subsequent FM state[48,54,55], depending on synthesis conditions. Sb doping also provides control over the chemical potential. In the investigation of $Mn(Bi_{1-x}Sb_x)_2Te_4$, tuning the chemical potential near the charge neutrality point was found to be critical for observing the ideal bulk Weyl state and layer-dependent tunable Chern insulator states [42,56]. Therefore, fine-tuning the chemical potential in $Mn(Bi_{1-x}Sb_x)_4Te_7$, along with its adjustable magnetism, presents a promising avenue for the discovery of novel topological quantum states. This potential motivates a systematic exploration of $Mn(Bi_{1-x}Sb_x)_4Te_7$ by precisely tuning its chemical potential and magnetic properties.

As shown below, $Mn(Bi_{1-x}Sb_x)_4Te_7$ exhibits a complex magnetic phase diagram. While $MnBi_4Te_7$ undergoes an AFM transition around 13 K [44], doping with 15–20% Sb stabilizes the interlayer FM coupling, leading to an FM transition at approximately 13.5 K [55]. However, when the chemical potential is tuned near the charge neutrality point (around $x \sim 0.27$), the system reverts to an AFM phase, which then transitions back to an FM phase at lower temperatures. This AFM-to-FM transition induces an electronic phase transition, resulting in a unique FM WSM. Our experiments not only detected the key transport signature of the WSM – chiral anomaly, but also revealed a three-dimensional (3D) quantum Hall effect (QHE) at high magnetic fields. Moreover, strong quantum oscillations were observed in samples with $x \approx 0.27$, with temperature dependence suggesting that the Weyl state emerges through a Fermi surface (FS) reconstruction driven by FM ordering. To further support our interpretation of the Weyl state, we combined the DFT calculations with the analytical model analysis to reveal that the experimentally detected Weyl state evolves from a parent nodal ring state in the $k \cdot p$ type of effective model in the presence of the mirror-z symmetry. Higher-order $k$-terms break the mirror-$z$ symmetry and split the nodal rings into two sets of

Weyl nodes, each containing six nodes connected by three-fold rotation and inversion symmetry. These Weyl nodes exhibit type-II characteristics and are highly anisotropic. In particular, the Fermi velocity along the *z*-axis is almost an order of magnitude smaller than that in the in-plane directions. This reduction of Fermi velocity along the *z*-axis makes the z-directional position of the Weyl nodes and the anomalous Hall conductivity highly sensitive to higher-order *k*-terms. The Fermi velocity reduction along the *z*-axis may also account for the observed 3D QHE in our experiments. These findings highlight the rich array of quantum states accessible through precise chemical manipulation of the Mn(Bi$_{1-x}$Sb$_x$)$_4$Te$_7$ system.

## II. RESULTS AND DISCUSSIONS

**Coupled evolution of magnetic and electronic properties**

To finely tune the chemical potential, we synthesized a series of Mn(Bi$_{1-x}$Sb$_x$)$_4$Te$_7$ samples with Sb content (*x*) ranging from 0 to 0.55 (refer to the Methods Section for details on sample synthesis and characterizations). Comprehensive magnetic and transport measurements were performed on these samples, revealing distinct coupled evolution of electronic and magnetic properties. Figure 1(a) illustrates the variation in carrier type, carrier density, and mobility as a function of Sb concentration *x* in Mn(Bi$_{1-x}$Sb$_x$)$_4$Te$_7$. Similar to Mn(Bi$_{1-x}$Sb$_x$)$_2$Te$_4$, Mn(Bi$_{1-x}$Sb$_x$)$_4$Te$_7$ undergoes a carrier type transition from electron to hole at approximately *x* = 0.27. The carrier density reaches its minimum near this composition, where it is reduced by two orders of magnitude compared to pristine MnBi$_4$Te$_7$. This significant reduction suggests that the chemical potential approaches the charge neutrality point at this composition. This conclusion is further confirmed by angle-resolved photoemission spectroscopy (ARPES) and scanning tunneling spectroscopy (STS) measurements, as discussed below. Notably, carrier mobility peaks around this critical composition. It is important to note, however, that Sb doping is not the only factor influencing carrier density and mobility; variations in growth methods and heat treatments can also have a significant impact on magnetic and electronic properties. This is further discussed in Supplementary Note 1 [57], which also explains why our magnetic phase diagram in Fig. 1(b) differs from previously reported results [48]. All samples presented in Figs. 1(a) and 1(b) were synthesized using a similar method as described in the Methods Section.

In addition to tuning the chemical potential, Sb doping at the Bi site also adjusts the interlayer magnetic coupling in Mn(Bi$_{1-x}$Sb$_x$)$_4$Te$_7$, as depicted in Fig. 1(b). The blue and red data points represent the AFM (Néel temperature, $T_N$) and FM (Curie temperature, $T_C$) transition temperatures, respectively. Pristine MnBi$_4$Te$_7$ exhibits interlayer AFM coupling below $T_N$~13 K. However, with 10% Sb doping, the interlayer coupling shifts from AFM to FM at approximately 9 K. As the Sb concentration increases to 15-20% or above 45%, the system exhibits only an FM transition below $T_C$ (~13.5 K) [55], indicating that higher Sb levels favor interlayer FM coupling. This trend of AFM-to-FM transition, driven by Sb doping, is attributed to an increase in Mn antisite defects at Sb sites, which promotes interlayer FM coupling [48,55]. Interestingly, this trend reverses as $x$ is in the range of $0.2 < x < 0.45$, where the system first undergoes an AFM transition before transforming to a FM state at even lower temperature. Notably, while $T_N$ shows small variation within this composition range, $T_C$ is highly dependent on $x$, with a minimum $T_C$ (~7.5 K) observed near $x \approx 0.3$. This result implies that the FM instability depends not only on Mn antisite defects but also on carrier density.

The coupled evolutions of the magnetic phase and carrier density, as shown in Figs. 1(a) and 1(b), clearly indicate that the lower carrier density near $x \sim 0.3$ weakens the interlayer FM coupling, leading to the reemergence of the AFM phase within the $0.2 < x < 0.45$ range. Among the various mechanisms supporting the FM phase, the Ruderman–Kittel–Kasuya–Yosida (RKKY) interaction, which operates through itinerant electrons, is known to be highly sensitive to the changes in carrier density [58,59]. Our observation of the strong dependence of $T_c$ on carrier density within the $0.2 < x < 0.45$ range suggests that the RKKY interaction plays a key role in mediating interlayer FM coupling in Mn(Bi$_{1-x}$Sb$_x$)$_4$Te$_7$. To further verify this interpretation, we prepared another sample with $x = 0.27$, which underwent additional heat treatment and exhibited a carrier density an order of magnitude higher than that of the $x = 0.27$ sample shown in Fig. 1(a). This sample did not display an AFM-to-FM transition; instead, it showed only a single FM transition, similar to those observed in samples with $0.15 \leqslant x \leqslant 0.2$ (see Supplementary Note 1 [57]). This strongly supports our argument that the RKKY interaction plays a critical role in stabilizing the FM phase in $Mn(Bi_{1-x}Sb_x)_4Te_7$.

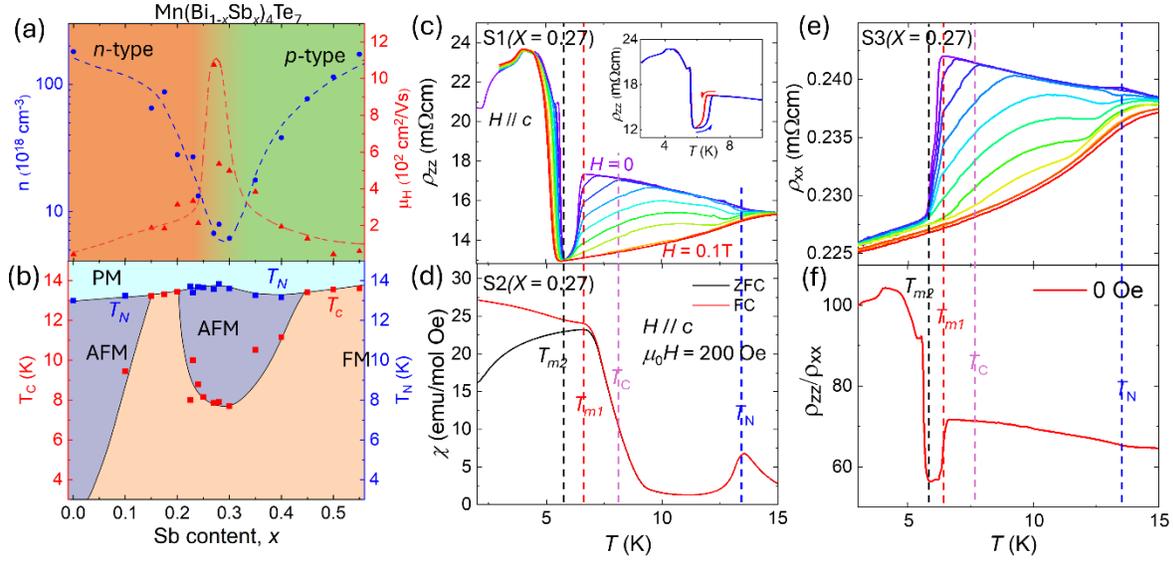

**FIG. 1**. Coupled evolution of electronic and magnetic properties for $Mn(Bi_{1-x}Sb_x)_4Te_7$ system. (a) Carrier density and mobility as a function of Sb doping concentration $x$. (b) Magnetic phase diagram. AFM and FM represent antiferromagnetic and ferromagnetic phase respectively; the blue and red data points denote the Neel ($T_N$) and Curie temperature ($T_c$) respectively. (c) Temperature dependent out-of-plane ($c$-axis) resistivity $\rho_{zz}$ at various magnetic fields measured on a $x = 0.27$ sample (S1). The field varies from 0 to 1000 Oe with a 100 Oe increment. The inset shows the zoomed in data at zero field. (d) Magnetic susceptibility as a function of temperature for a $x = 0.27$ sample (S2) measured with zero-field cooling (ZFC) and field cooling (FC) histories under a field of 200 Oe applied along the $c$-axis. The blue, purple red and black dashed lines in (c) and (d) mark several characteristic temperatures, i.e. $T_N$, $T_c$, $T_{m1}$ & $T_{m2}$. $T_{m1}$ & $T_{m2}$ mark the temperature window where thermal hysteresis occurs to the AFM-to-FM transition [see the inset to (c)]. (e) Temperature dependent in-plane resistivity $\rho_{xx}$ at various magnetic fields measured on a $x = 0.27$ sample (S3). The field varies from 0 to 1000 Oe with a 100 Oe increment (with similar color scheme labeling the fields of the curves as in (c)). (f) Temperature dependence of the $\rho_{zz}/\rho_{xx}$ ratio at zero field.

**Electronic phase transition driven by the AFM-to-FM transition near $x \sim 0.27$**

Another significant manifestation of coupling between electronic and magnetic properties in $Mn(Bi_{1-x}Sb_x)_4Te_7$ is the electronic phase transition driven by the AFM-to-FM transition in samples where the chemical potential is near the charge neutrality point ($x \sim 0.27$). This transition is evident in the electrical transport measurements. Fig. 1(c) shows the out-of-plane resistivity $\rho_{zz}$ as a function of temperature for a representative sample of $x = 0.27$ (denoted S1). Since $\rho_{zz}$ is sensitive to interlayer spin scattering, it is expected to show distinct features at magnetic transitions. As previously noted, the $x = 0.27$ sample shows an AFM state below $T_N$ (~ 13 K), which transitions to a FM state below $T_c$. The magnetic susceptibility $\chi$ data in Fig. 1(d) confirms these two successive magnetic transitions, with $T_N$ and $T_c$ being ~ 13.5 K and 7.8 K. From Fig. 1(c), it is clear that $\rho_{zz}$ exhibits noticeable anomalies at both $T_N$ and $T_c$: an upturn at $T_N$ and downturn at $T_c$, as indicated by the vertical dashed blue ($T_N$) and purple ($T_c$) lines (e.g. see the data taken under the field of 200 Oe, the same field used for the $\chi$ measurements shown in Fig. 1(d). Note that Tc shifts to higher temperatures with increasing applied field). The upturn of $\rho_{zz}$ below $T_N$ is attributed to increased interlayer spin scattering caused by the A-type AFM ordering [44], while the downturn below $T_c$ is due to the suppression of interlayer spin scattering resulting from FM ordering.

In addition to these two anomalies, $\rho_{zz}$ also displays two unusual anomalies: a steep drop near 6.3 K (denoted as $T_{m1}$) and a sharp upturn near 5.6 K (denoted as $T_{m2}$). The gradual decrease of $\rho_{zz}$ between $T_c$ to $T_{m1}$ can be attributed to the progressive formation of FM domain in the AFM background, while the steep drop at $T_{m1}$ likely results from the percolation of the FM phase. Given that the FM phase has much lower resistivity than the AFM phase, the percolation of the FM phase naturally leads to a sharp resistivity drop. This interpretation is supported by the observation that, when the AFM phase is polarized to a FM phase under a field exceeding 800 Oe, the steep drop at $T_{m1}$ disappears. Moreover, upward and downward temperature sweep measurements of $\rho_{zz}$ reveal thermal hysteresis within the $T_{m2} < T < T_{m1}$ temperature regime (see the inset to Fig. 1(c)), suggesting that the AFM-to-FM transition is first-order in nature, with notable AFM and FM phase coexistence and FM phase percolation occurring within this temperature window. This conclusion aligns with the magnetic susceptibility data which shows pronounced bifurcation between the zero-field-cooling (ZFC) and field cooling (FC) histories starting from $T_{m1}$ (Fig. 1(d)).

Given that the magnetic phase below $T_{m2}$ is expected to be purely FM, the sharp upturn in $\rho_{zz}$ below $T_{m2}$ is quite surprising. Typically, an AFM-to-FM transition is associated with reduced resistivity due to suppressed spin scattering, as observed in the temperature range $T_{m2} < T < T_c$. One possible explanation for the sharp increase of $\rho_{zz}$ below $T_{m2}$ is that the interlayer FM coupling causes significant spin splitting of electronic bands, leading to an electronic structure transition below $T_{m2}$. In previous studies on Mn(Bi$_{1-x}$Sb$_x$)$_2$Te$_4$, it was found that the AFM-to-FM transition induced by a magnetic field drives a change from an AFM topological insulator (TI) to an ideal WSM, a transition observed only when the chemical potential is close to the charge neutrality point [42,43]. Given that our $x = 0.27$ sample has a chemical potential near this point, as demonstrated by ARPES and STS experiment (see below), it is plausible that an electronic phase transition occurs as the AFM phase spontaneously evolves into an FM phase below $T_{m2}$. Our $\rho_{zz}$ data (Fig. 1(c)), along with in-plane resistivity $\rho_{xx}$ measurements (Fig. 1(e)), support this hypothesis. We observe a sharp drop in $\rho_{xx}$ at $T_{m1}$ and a lower resistivity state with a metallic temperature dependence below $T_{m2}$, contrasted with the high resistivity state with the non-metallic temperature dependence in the AFM phase (Fig. 1(e)). Additionally, the electronic state below $T_{m2}$ becomes much more anisotropic, as indicated by the striking increase of the $\rho_{zz}/\rho_{xx}$ ratio below $T_{m2}$ (Fig. 1(f)). The value of $\rho_{zz}/\rho_{xx}$ rises from ~65 – 72 in the AFM state between $T_N$ and $T_{m1}$ up to ~91-105 in the FM state below $T_{m2}$. Although fields above 800 Oe can polarize the AFM phase between $T_N$ and $T_{m2}$ into an FM phase (Supplementary Fig. 1(d) [57]), the transition to a highly anisotropic state with larger $\rho_{zz}$ occurs only at $T_{m2}$, not at $T_N$ (see data taken at 0.08 T and 0.1 T in Fig. 1(c)). These results suggest that only the uniform spontaneous interlayer FM coupling below $T_{m2}$ is capable of driving a significant electronic transition.

We investigated the electronic band structure of Mn(Bi$_{1-x}$Sb$_x$)$_4$Te$_7$ ($x = 0.27$) using ARPES and scanning tunneling microscopy and spectroscopy (STM/STS). Figure 2(a) shows the constant energy contours at various binding energies ($E_B$) in the $k_x$-$k_y$ plane (left) and the band dispersion along the $\Gamma$-$K$ direction (right) at 6 K, near $T_{m_1}$. The data indicates that the sample is lightly hole-doped, with the valence band top crossing the Fermi level only near the Γ point. This suggests that ~27% Sb doping significantly lowers the chemical potential compared to pristine MnBi$_4$Te$_7$, which is heavily electron-doped [44]. From the point-like Fermi surface observed at $E_B = 0$ (left, Fig. 2(a)), the Fermi wave vector $k_F$ is estimated to be 0.05±0.02 Å$^{-1}$, which is very small compared to the Brillouin zone size. This suggests that the chemical potential is close to the charge neutrality point, consistent with the low carrier density extracted from transport measurements, as discussed above.

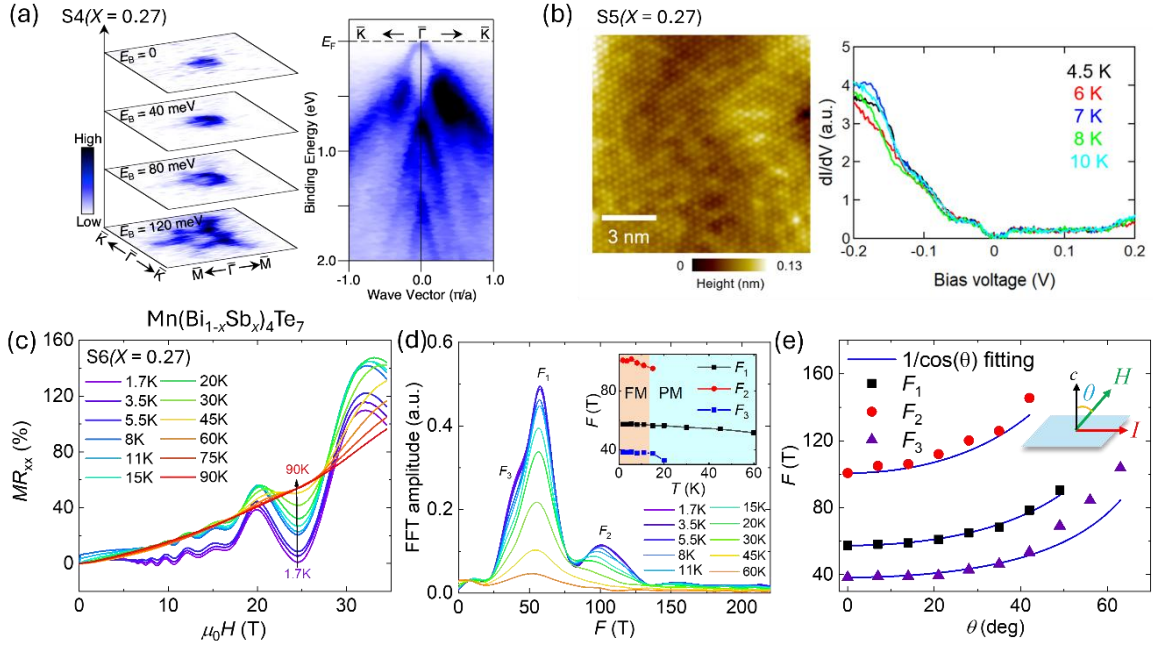

**FIG. 2**. Fermi surface characteristics of Mn(Bi$_{1-x}$Sb$_x$)$_4$Te$_7$ ($x$ = 0.27) probed by angle-resolved photoemission spectroscopy (ARPES), scanning tunneling microscopy/spectroscopy (STM/STS) and Shubnikov-de Haas (SdH) oscillations. (a) Constant energy contours at various binding energies ($E_B$) on the $k_x$-$k_y$ plane (left) and band dispersion along the $\overline{\Gamma K}$ cut (right) probed by ARPES at 6 K with the photon energy of $h\nu$ = 90 eV. (b) STM image (over 12×12 nm$^2$) of the (Bi,Sb)$_2$Te$_3$ termination surface taken at 4.5 K (left) and the STS data measured on the (Bi,Sb)$_2$Te$_3$ termination surface at various temperatures (right). $V$ = −0.5 V, $I$ = 0.5 nA. (c) In-plane magnetoresistivity, defined as MR$_{xx}$ = [ρ$_{xx}$(H)-ρ$_{xx}$(0)]/ρ$_{xx}$(0), vs. magnetic fields (applied along the out-of-plane direction) at various temperatures. (d) FFT spectra of -d$^2$ρ$_{xx}$/d($\mu_0 H$)$^2$ at various temperatures. The inset shows the temperature dependence of the quantum oscillation frequency extracted from the FFT spectra. (e) Angular dependence of SdH oscillation frequencies $F_1$, $F_2$ and $F_3$. The solid lines are the fits to 1/cos(θ). The inset to (c) shows the experimental setup for angle-dependent MR measurements.

Although the ARPES sample is lightly hole-doped, preventing the observation of its topological surface state (TSS), our STS measurements revealed the TSS. The right panel of Fig. 2(b) shows STS data measured on the (Bi,Sb)$_2$Te$_3$ termination surface (left, Fig. 2(b)) at various temperatures. This dataset indicates that the Fermi level in this flat region (~12×12 nm$^2$, left, Fig. 2(b)) is at the charge neutrality point and reveals a homogeneous, temperature-independent gap (up to 10 K) of ~40 meV. Based on previous ARPES and STM/STS studies on MnBi$_4$Te$_7$ [60,61], this surface gap opening can be attributed to the hybridization between the TSS and the bulk band. On the Mn(Bi,Sb)$_2$Te$_4$ termination surface, we observed a V-shaped tunneling conductance curve (dI/dV vs. bias voltage) (see supplementary Note 2 [57]), consistent with the gapless surface state previously probed on the MnBi$_2$Te$_4$ termination surface of MnBi$_4$Te$_7$ in ARPES and STS measurements [60,61].

To further investigate the electronic structure changes driven by spontaneous interlayer FM coupling, we performed magneto-transport measurements on a representative sample with $x = 0.27$ (denoted as S6) under high magnetic fields (up to 35 T) at the National High Magnetic Field Laboratory (Tallassee, Florida). The high carrier mobility of this sample (Fig. 1(a)) allowed us to observe strong Shubnikov–de Haas (SdH) oscillations (Fig. 2(c)), providing an opportunity to examine whether FM ordering induces FS reconstruction due to spin splitting. From temperature-dependent SdH oscillation measurements (Fig. 2(c)), we found clear evidence of FS reconstruction induced by FM ordering. Since the SdH oscillation frequency is proportional to the extremal cross-section area of the FS, any electronic structure transition caused by FM ordering would alter the FS morphology, leading to changes in the quantum oscillation frequency across the FM transition. To test this, we conducted fast Fourier transform (FFT) analysis on the second derivative of the magnetoresistivity data (Fig. 2(c)), with the results presented in Fig. 2(d). A single oscillation frequency ($F_1$) was observed in the paramagnetic (PM) state above the critical temperature ($T_c$ = 13.5 K), while three frequencies ($F_1$, $F_2$, and $F_3$) emerged in the FM state below $T_c$, with $F_1$ being the dominant component. Notably, since the AFM phase is polarized into the FM state above 800 Oe, all SdH oscillations in Fig. 2(c) correspond to the FM phase. At the base temperature of 1.7 K, the frequencies $F_1$, $F_2$, and $F_3$ were 57.4 T, 100.7 T, and 38.4 T, respectively. While F$_2$ is not far from the second harmonic of $F_1$, its remarkable enhancement below $T_C$ indicates it is associated with the FM transition. From $F_1$, the extremal cross-sectional area of the FS was estimated to be 0.5 nm$^{-2}$, which corresponds to approximately 0.23% of the total area of the first Brillouin zone with $k_F$ being 0.04 Å$^{-1}$, consistent with the small $k_F$ value (~ 0.05 Å$^{-1}$) estimated from ARPES. The appearance of additional frequency components $F_2$ and $F_3$ below $T_c$ indicates FS reconstruction across the FM transition and aligns with the expectation that type-II Weyl state is characterized by coexistence of electron and hole pockets. Our DFT calculations based on the composition Mn(Bi$_{0.75}$Sb$_{0.25}$)$_4$Te$_7$ suggest that the band structure change driven by the AFM-to-FM transition leads the density of state (DOS) at the Fermi level to decrease from 42.91 states/eV (AFM) state to 34.24 states/eV (FM) (see Supplementary Note 3 [57]).

Furthermore, we observed that $F_1$, $F_2$, and $F_3$ varied with temperature (inset of Fig. 2(d)). Although $F_1$ did not show a drastic change at $T_c$, its temperature dependence was significant, as seen by the shift of the oscillation peak near 32 T in Fig. 2(c). This temperature-dependent quantum oscillation frequency likely results from strong coupling between the electronic band structure and magnetism, as previously discussed in studies on Mn(Bi$_{1-x}$Sb$_x$)$_2$Te$_4$ [42]. Lower temperatures suppress thermal fluctuations, enhancing ordered magnetic moments, which should increase band splitting and result in a progressive FS variation. Interestingly, this temperature dependence of frequency extends above $T_c$, likely due to short-range FM order persisting in the PM state. Additionally, we also measured the angular dependence of $F_1$, $F_2$, and $F_3$ (Fig. 2(e)) and found that $F_1(\theta)$ and $F_2(\theta)$ follow a $1/\cos(\theta)$ relationship, while $F_3$ deviates from this trend above 40°. These findings suggest that the FM state in the $x = 0.27$ sample primarily exhibits a quasi-two-dimensional (2D) electronic structure, consistent with the increased electronic anisotropy revealed in resistivity measurements (Fig. 1(f)) as discussed above.

**Transport signatures of the FM Weyl semimetal state**

The Fermi surface reconstruction associated with the AFM-to-FM transition suggests the potential emergence of a new topological state. Early DFT band structure calculations propose that MnBi$_4$Te$_7$ should host a WSM state if its interlayer magnetic coupling switches from AFM to FM [48,49]. While previous studies have shown that pristine MnBi$_4$Te$_7$ prepared via the melt growth method shows FM-like behavior below 5K [49], and Sb-doped MnBi$_4$Te$_7$, either undergoes a single FM transition or transition first to an AFM state and then to an FM state [48,54,55], no experimental evidence for a FM WSM state has been reported for MnBi$_4$Te$_7$ or Mn(Bi$_{1-x}$Sb$_x$)$_4$Te$_7$. The likely reason is that the Weyl nodes in previously studied samples are not located near the Fermi level. However, through our systematic studies, we find that the Fermi surface reconstruction observed in the $x \sim 0.27$ sample is indeed associated with a WSM state. The key transport signature of this WSM state - chiral anomaly effect was revealed in magnetotransport measurements of the $x \sim 0.27$ sample.

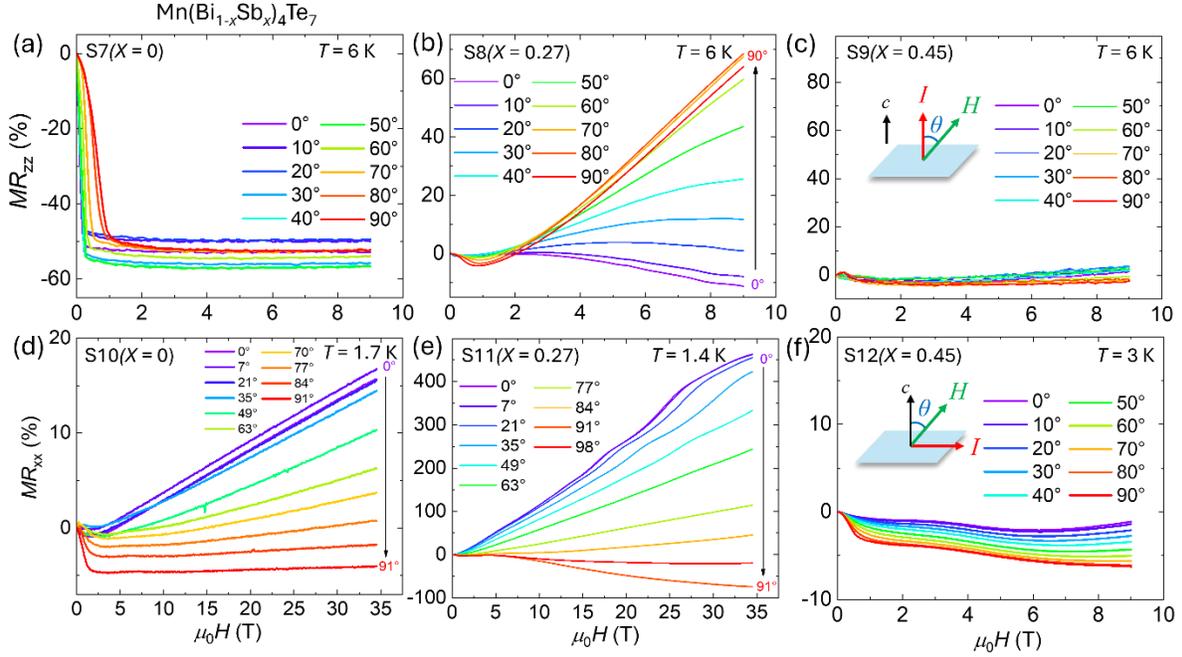

**FIG. 3**. Magnetoresistivity (MR) of Mn(Bi$_{1-x}$Sb$_x$)$_4$Te$_7$ under various field orientations at 1.4 - 6 K. The $c$-axis magnetoresistivity $MR_{zz} = [\rho_{zz}(H) - \rho_{zz}(0)]/\rho_{zz}(0)$ measured under various field orientations for $x = 0$ (a), $x = 0.27$ (b), and $x = 0.45$ (c). The inset to (c) shows the experimental configuration for MR$_{zz}$ measurements under a given field orientation angle $\theta$. The in-plane magnetoresistivity $MR_{xx} = [\rho_{xx}(H) - \rho_{xx}(0)]/\rho_{xx}(0)\}$ measured up to 35 T under various field orientations for $x = 0$ (d), $x = 0.27$ (e), and $x = 0.6$ (f). The inset to (f) shows the experimental configuration for $MR_{xx}$ measurements under a given field orientation angle $\theta$.

The chiral anomaly is a hallmark of WSMs and is characterized by negative longitudinal magnetoresistance (MR) when electric and magnetic fields are applied in parallel. This arises from the topological current $j_c$, induced by the chirality imbalance of Weyl fermions, where $j_c \propto \vec{E}\cdot\vec{B}\,\tau_{int}$, with $\vec{E}$ and $\vec{B}$ representing the electric and magnetic fields, respectively, and $\tau_{int}$ denoting the inter-Weyl node relaxation time. As a result, the chiral anomaly effect depends sensitively on the magnetic field orientation relative to the current direction. The negative MR induced by the chiral anomaly reaches its maximal magnitude when $B \parallel I$ but gradually diminishes as the magnetic field rotates away from the current. This is exactly what we observed in the $x = 0.27$ samples. Figure 3(b) shows the c-axis magnetoresistivity $MR_{zz} = [\rho_{zz}(H) - \rho_{zz}(0)]/\rho_{zz}(0)$ of a representative $x = 0.27$ sample (S8), measured under various magnetic field orientations at 6 K (see the inset in Fig. 3(c)). $MR_{zz}$ is significantly negative above 6 T (e.g., ~ -11% at 9 T) when $\theta = 0°$, where $B \parallel I$, but evolves into positive values across the entire measured field range when $\theta > 20°$, showing large positive values (e.g. ~70% at 9 T) as $\theta$ approaches 90°. This negative-to-positive evolution of $MR_{zz}$ with $\theta$ contrasts sharply with the $MR_{zz}$ data of the heavily electron-doped sample (e.g., pristine MnBi$_4$Te$_7$, Fig. 3(a)) and the heavily hole-doped sample (e.g., $x = 0.45$, Fig. 3(c)).

Although MnBi$_4$Te$_7$ is AFM, it can be driven into a FM state when the magnetic field along the c-axis exceeds 0.1 T. As shown in Fig. 3(a), this field-induced AFM-to-FM transition leads to a step-like drop in $MR_{zz}$ (~ -50% for $\theta = 0°$), after which $MR_{zz}$ remains constant in the FM phase with further increases in the magnetic field. This behavior can be interpreted as a spin-valve effect, where the transition from interlayer antiparallel spin alignment to parallel spin alignment significantly reduces spin scattering, causing a sharp drop in $MR_{zz}$. Magnetic field orientation has only a minor effect on the spin-valve behavior. The $x = 0.45$ sample shows only an FM transition at $T_c \sim 13.5$ K, with very small and nearly $\theta$-independent $MR_{zz}$. The sharp contrast in $MR_{zz}$ data between the $x = 0.27$ sample and the $x = 0$ and $x = 0.45$ samples rules out other origins of negative longitudinal MR for the $x = 0.27$ sample, such as suppression of spin scattering or current jetting, and suggests that the c-axis magnetotransport is influenced by the chiral anomaly effect. This is further supported by $MR_{zz}$ measurements on the $x = 0.23$, 0.28, and 0.30 samples (see Supplementary Fig. 8 [57]). Although all these samples share similar magnetic transitions, as shown in Fig. 1(b), only the

$x$ = 0.28 and 0.30 samples show chiral anomaly behavior in $MR_{zz}$, while the $x$ = 0.23 sample shows a spin-valve effect. This difference arises because the $x$ = 0.28 and 0.30 samples, like the $x$ = 0.27 sample, have chemical potentials close to the charge neutrality point, as indicated by their measured carrier densities, whereas the chemical potential of the $x$ = 0.23 sample is still not close to the charge neutrality point (see Fig. 1(a)).

Unlike the ideal Weyl state in the FM phase of Mn(Bi$_{1-x}$Sb$_x$)$_2$Te$_4$ ($x$ = 0.26), which features only one pair of Weyl nodes along the $k_z$-axis, the WSM state of Mn(Bi$_{1-x}$Sb$_x$)$_4$Te$_7$ ($x \sim$ 0.27) contains six pairs of Weyl nodes, positioned at different momentum points in the $k_x$-$k_y$ plane as discussed below. Therefore, the chiral anomaly effect is also expected in the in-plane magnetotransport. To verify this, we measured the in-plane magnetoresistivity $MR_{xx}$ for the $x$ = 0, 0.27 and 0.45 samples. We observed the chiral anomaly effect only in the $x$ = 0.27 sample, where $MR_{xx}$ becomes significantly negative above 10 T for $\theta$ = 90°, where $B \parallel I$ (see the inset in Fig. 3(f)), without saturation, reaching ~ -75% at 35 T. However, as the field is rotated toward the c-axis, $MR_{xx}$ becomes positive and reaches large positive values at high fields for $B \perp I$ (e.g., ~ 450% at 35 T). In contrast, the $x$ = 0 sample shows a spin-valve effect under $B \parallel I$ (Fig. 3(d)), and the $x$ = 0.45 sample displays very small, weakly $\theta$-dependent negative MR caused by the suppression of spin scattering (Fig. 3(f)). This contrast again indicates that the Weyl fermion transport is observable only when the Weyl nodes are tuned to be close to the Fermi level at $x \sim$ 0.27.

In general, a large anomalous Hall effect (AHE) is expected for FM WSMs due to the diverging Berry curvature at the Weyl nodes and the breaking of time reversal symmetry. This phenomenon has been well demonstrated in previously reported FM WSMs such as Co$_3$Sn$_2$S$_2$ and Co$_2$MnGa [62,63]. However, we did not observe a significant AHE in Mn(Bi$_{1-x}$Sb$_x$)$_4$Te$_7$ with the maximal anomalous Hall conductivity being around 13 $\Omega^{-1}$cm$^{-1}$ for $x \sim$ 0. 27 (see supplementary Note 4 [57]). As explained below, the small AHE is because the Weyl nodes originate from the splitting of the nodal rings and requires the high-order momentum terms, which are expected to be small. Indeed, the calculated anomalous Hall conductivity is small, consistent with the experimental anomalous Hall conductivity (see supplementary Note 4 and Fig. 5f [57]).

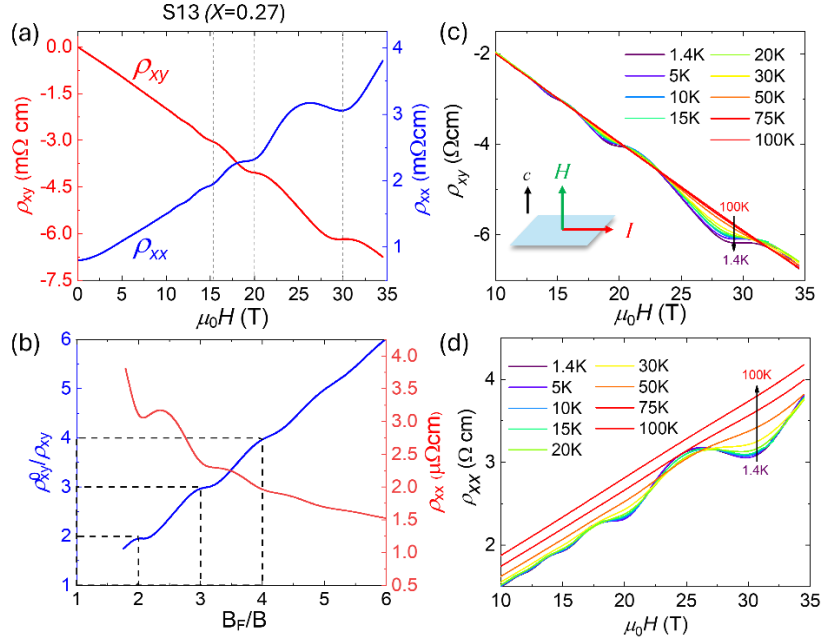

**FIG. 4.** Bulk quantum Hall effect in Mn(Bi$_{1-x}$Sb$_x$)$_4$Te$_7$ ($x$ = 0.27). (a) In-plane longitudinal resistivity $\rho_{xx}$ (blue) and Hall resistivity $\rho_{xy}$ (red) as functions of the external magnetic field applied along the $c$-axis at 1.4K for a representative $x$ = 0. 27 sample (S13). (b) Normalized inverse Hall resistivity ($\rho_{xy}^0/\rho_{xy}$) and $\rho_{xx}$ as a function of $B_F/B$. $\rho_{xy}^0$ is defined as the inverse of the step size between nearest plateaus of $1/\rho_{xy}$ (see supplementary Fig. 9 [57]), and $B_F$ is the SdH oscillation frequency. (c) and (d) Field dependence of $\rho_{xy}$ and $\rho_{xx}$ (d) at various temperatures. The Inset to (c) shows the experimental configuration for $\rho_{xx}$ and $\rho_{xy}$ measurements.

**Bulk quantum Hall effect of Mn(Bi$_{1-x}$Sb$_x$)$_4$Te$_7$ ($x \sim 0.27$)**

In addition to the chiral anomaly effect, Mn(Bi$_{1-x}$Sb$_x$)$_4$Te$_7$ ($x \sim 0.27$) also exhibits another distinct quantum transport phenomenon – bulk quantum Hall effect (QHE). Figure 4(a) presents the longitudinal resistivity ($\rho_{xx}$) and Hall resistivity ($\rho_{xy}$) data of a representative $x$ = 0.27 sample (S13), measured as a function of magnetic field at 1.4 K with the field applied along the $c$-axis (see the inset to Fig. 4c). $\rho_{xx}$ exhibits remarkable SdH oscillations with the quantum oscillation frequency $B_F$ = 61.8 T. As $\rho_{xx}$ shows minima at ~15 T, 20 T, and 30 T, $\rho_{xy}$ display well developed plateau behavior, implying the presence of bulk QHE. From the normalization of $1/\rho_{xy}$ by the inverse of the step height between the $\rho_{xy}$ plateaus at

20 T and 30T (denoted as $1/\rho_{xy}^0$, see the supplementary Fig. 9 [57]), we find solid evidence of QHE. Figure 4(b) shows $\rho_{xy}^0/\rho_{xy}$ as a function of $B_F/B$. $\rho_{xy}^0/\rho_{xy}$ is clearly quantized to integer numbers 2, 3, and 4 at the $\rho_{xy}$ plateaus, which, respectively, corresponds to the integer normalized filling factor $B_F/B$ of 2, 3 and 4 at the $\rho_{xx}$ minima. All these features can be attributed to QHE. Both the $\rho_{xy}$ plateaus and $\rho_{xx}$ oscillations sustain up to ~30 K and then diminish as the temperature is increased above 30 K, as shown in Figs. 4(c) and 4(d).

While the quantum Hall effect (QHE) is typically observed in two-dimensional (2D) electron systems, it has also been reported in several three-dimensional (3D) materials, including electron-doped $Bi_2Se_3$ [64], $EuMnBi_2$ [65], $BaMnSb_2$ [66], $ZrTe_5$ [67], $HfTe_5$ [68], and $Cd_3As_2$ [69]. These 3D manifestations of the QHE arise from distinct mechanisms. In $Bi_2Se_3$, $EuMnBi_2$, and $BaMnSb_2$ crystals, the QHE can be attributed to a stacked QHE mechanism, owing to their quasi-2D electronic structures and weak interlayer coupling. The 3D QHE observed in $ZrTe_5$ scales with the Fermi wavevector and has been interpreted as originating from a field-induced charge-density wave caused by electron interaction effects [67,70], though alternative mechanisms have also been proposed [67,71]. In $Cd_3As_2$, the QHE is ascribed to Weyl orbital physics [69].

Our observation of the QHE in $Mn(Bi_{1-x}Sb_x)_4Te_7$ ($x \sim 0.27$) is likely tied to its quasi-2D electronic band structure. This hypothesis aligns with the angular dependence of the quantum oscillation frequency shown in Fig. 2(e) and the high resistivity anisotropy ($\rho_{zz}/\rho_{xx}$) ratio shown in Fig. 1(f). Moreover, the coexistence of a chiral anomaly and Fermi surface reconstruction driven by FM ordering suggests that $Mn(Bi_{1-x}Sb_x)_4Te_7$ ($x \sim 0.27$) hosts a FM WSM state. The emergence of a robust QHE within this state further indicates that the WSM is highly anisotropic, exhibiting quasi-2D characteristics. Notably, none of the established WSMs have demonstrated 3D QHE. In $Mn(Bi_{1-x}Sb_x)_4Te_7$ ($x \sim 0.27$), the period of the QHE layer, approximately 5.0 nm (see Supplementary Note 5 [56]), is about twice the atomic lattice constant along the z-axis (2.37 nm [44]). This observation is not in line with the simplistic stacked QHE model and suggests a more complex origin. Thus, the 3D QHE observed in Mn(Bi$_{1-x}$Sb$_x$)$_4$Te$_7$ (x ~ 0.27) offers a unique platform to explore quantum Hall physics in the context of Weyl fermions.

**Theoretical modelling of anisotropic type-II Weyl nodes**

To understand the above experimental observations, we performed the DFT calculations of the electronic band structure of MnBi$_4$Te$_7$ in the FM phase and constructed a three-band model to identify the physical origin, the momentum space locations and Fermi velocities of the Weyl nodes.

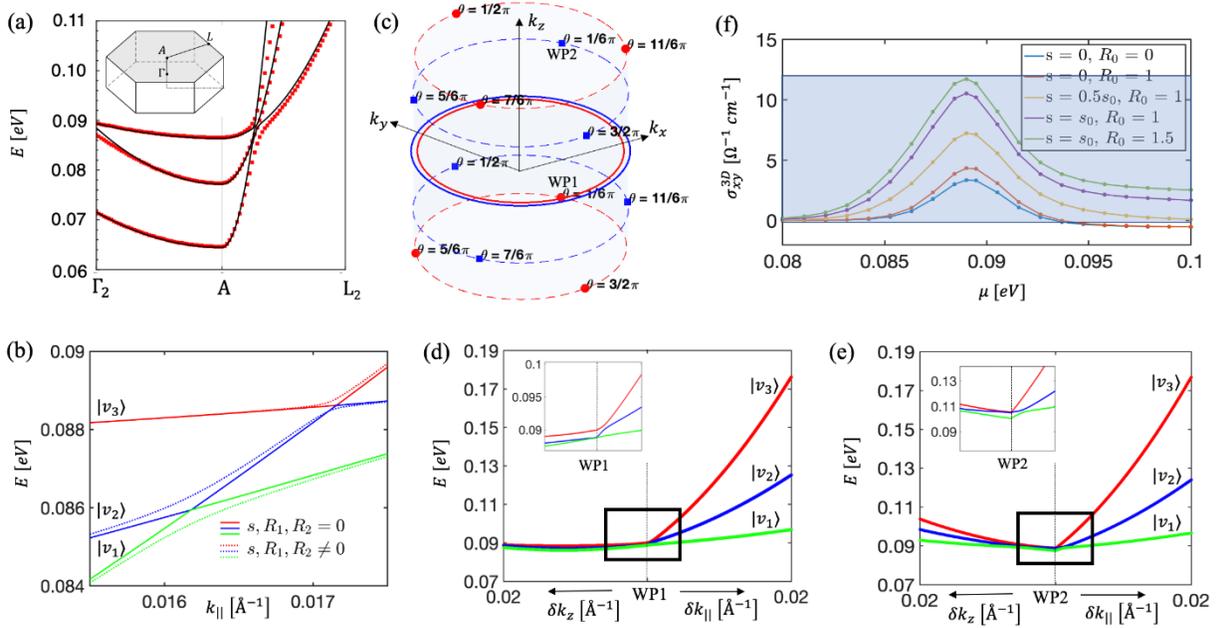

**FIG. 5**. Theoretical modelling of an anisotropic type-II Weyl state for MnBi$_4$Te$_7$. (a) The comparison between the DFT band structure (in red dots) and the energy dispersion of the 3-band model $H_{3b}$ (in black lines) along the $\Gamma_2 - A - L_2$ lines where $\Gamma_2 = (0,0,5/12)$, $A = (0,0,1/2)$, $L_2 = (0, 1/25, 1/2)$ in units of $\vec{b}_{1,2,3}$. (b) The energy dispersion of the 3-band model as a function of in-plane momenta $k_\parallel$ along $k_z = 0$ and $\theta = \pi/6$ when full rotation symmetry is present i.e., $R_1 = R_2 = s = 0$ (solid lines). The two band crossings are gapped (dashed lines) when full rotation is broken to 3-fold rotation $C_{3z}$. (c) When full rotation symmetry is present, the band crossings lead to 2 nodal rings at $k_3 = \pi$ ($k_z = 0$) with radii $k_\parallel^{red} = 0.01618$ Å$^{-1}$ and $k_\parallel^{blue} = 0.01716$ Å$^{-1}$, whereas when full rotation symmetry is broken, the nodal rings split into 12 Weyl nodes located at $k_z^{red} = \pm 0.0115, k_z^{blue} = \pm 0.007, k_\parallel^{red} = 0.01657, k_\parallel^{blue} = 0.01687$, in units of Å$^{-1}$. (d) The anisotropic energy dispersion at the Weyl node (WP1: $k_\parallel = 0.01657$ Å$^{-1}, \theta = \frac{\pi}{6}, k_z = -0.0115$ Å$^{-1}$) along the $k_z$ on the left and along $k_\parallel$ on the right. (e) The anisotropic energy dispersion at the Weyl node (WP2: $k_\parallel = 0.01687$ Å$^{-1}$, $\theta = \frac{\pi}{6}$, $k_z = 0.007$ Å$^{-1}$) along the $k_z$ on the left and along $k_\parallel$ on the right. The insets in (d) and (e) zoom-in of the black boxed regions of the energy dispersion. (f) The 3D Anomalous Hall conductivity as a function of chemical potential for different strengths of the full rotation breaking term ($s, R_1, R_2$) at kT = 1 meV. The shaded region corresponds to the range of experimental observed values. We define $s_0 = 1.38879$ eV Å$^2$ and dimensionless parameter $R_0$ such that $R_1 = 15.769 R_0$ eV Å$^3$ and $R_2 = 5.88178 R_0$ eV Å$^3$.

The DFT band structure in the full energy range is shown in Supplementary Note 6 [57]. As we are interested in the case with the Fermi energy close to the charge neutrality point, we focus on the lowest energy conduction bands, of which the DFT bands are shown in Fig. 5 (a) with red dots and can be fitted well by the tight-binding model constructed from the Wannier function method (See Supplementary Note 6 [57] for the details of the construction of Wannier function model). We project the tight-binding model into the subspace spanned by the three lowest energy conduction bands and obtain a three-band model $H_{3b}$, the detailed form of which is described in **Methods Sec. B**. As this model is expanded around A point, the $k_z = 0$ plane corresponds to the A-H-L plane in the Brillouin zone, which can be found in Supplementary Note 7 [57]. The energy dispersion of $H_{3b}$, as shown in black lines, can reproduce the DFT bands depicted by red dots in Fig. 5(a). The full Hamiltonian $H_{3b}$ can also be constructed based on the magnetic point group -3m', generated by three symmetry operators, inversion $\hat{P}$, three-fold rotation $\hat{C}_{3z}$ and the anti-unitary symmetry $\hat{C}_{2x}\hat{T}$, which combines the two-fold rotation $\hat{C}_{2x}$ and time reversal $\hat{T}$. The details can be found in Supplementary Note 6 [57]. The basis wavefunctions of $H_{3b}$ can be labelled by the eigen-values of $\hat{P}$ and $\hat{C}_{3z}$, denoted as $\{|v_1\rangle = |\Psi_-^-\rangle, |v_2\rangle = |\Psi_+^-\rangle, |v_3\rangle = |\Psi_-^+\rangle\}$ where $\hat{P}|\Psi_\beta^\alpha\rangle = \alpha|\Psi_\beta^\alpha\rangle$, $\hat{C}_{3z}|\Psi_\beta^\alpha\rangle = e^{i\beta\frac{\pi}{3}}|\Psi_\beta^\alpha\rangle$ in Fig. 5(b), with $\alpha, \beta = \pm$. The full Hamiltonian $H_{3b}$ can be decomposed into two parts, $H_{3b} = H_0 + H_1$ (See **Methods Sec. B**), in which $H_0$ has additional mirror symmetry $\hat{m}_z$ at the $k_z = 0$ plane. For an eigen-state, the eigen-value of $\hat{m}_z$ can be connected to the eigen-values of $\hat{P}$ and $\hat{C}_{3z}$ by $i\alpha\beta$. The existence of $\hat{m}_z$ guarantees two nodal rings at $k_z = 0$ in the energy dispersions of $H_0$, as labelled by the red and blue circles in Fig. 5(c). Here the red nodal ring comes from the band crossing between $|v_1\rangle$ and $|v_2\rangle$, while the blue nodal ring is from the band crossing between $|v_1\rangle$ and $|v_3\rangle$, as shown in Fig. 5(b). On the other hand, $H_1$ includes higher-order $k$ terms that break the $\hat{m}_z$ symmetry and thus splits both the red and blue nodal rings into two sets of Weyl nodes, denoted as WP1 and WP2, with each set containing six Weyl nodes that can be connected by $\hat{P}$ and $\hat{C}_{3z}$, as labelled by red dots and blue squares in Fig 5(c). We can further project the three-band model into the subspace spanned by the two bands for each Weyl node and determine the location of the Weyl node analytically (See the Supplementary Note 8 [57]). The Weyl node location along the $k_z$ direction sensitively depends on the parameter $s$ in $H_1$, while the location along the in-plane direction $k_\parallel$ is almost independent of the parameters in $H_1$. Figure 5(d-e) shows the anisotropic energy dispersion for WP1 and WP2 along the $k_\parallel$ and $k_z$ directions. As the conductivity of the whole system is mainly determined by carriers with the largest velocity, we extract the maximal Fermi velocity $v_z$ along the $k_z$ direction and $v_\parallel$ along the $k_\parallel$ direction of these type-II Weyl nodes, and find $|v_z| = 0.3149$ eV Å, $|v_\parallel| = 2.4442$ of eV Å at WP1 and $|v_z| = 0.1847$ eV Å, $|v_\parallel| = 1.4874$ eV Å at WP2. Thus, the maximal $|v_\parallel|$ is generally one order larger than $|v_z|$. Supplementary Fig. 13 in Supplementary Note 9 [57] further shows $v_z$ and $v_\parallel$ as a function of the parameter $s$, from which one can see the strong anisotropy of the Fermi velocity between the in-plane and out-of-plane directions generally exists, independent of the detailed parameter values. The small Fermi velocity $v_z$ suggests the quasi-2D nature of the Weyl electrons and allows for quasi-2D Landau levels with large density of states under magnetic fields [70,72], which is presumably responsible for the observation of 3D QHE in FM Mn(Bi$_{1-x}$Sb$_x$)$_4$Te$_7$. This also implies the sensitivity of the z-directional locations for Weyl nodes, and Supplementary Fig. 13(a) of Supplementary Note 9 [57] shows the dependence of the z-direction locations of Weyl nodes on the parameter $s$ in $H_1$. We further evaluate the Hall conductivity as a function of Fermi energy for different parameter sets of $H_1$ (Fig. 5(f)), which generally exhibits a peak value when the Fermi energy is close to the Weyl points (See Supplementary Note 7 and 8 [57]). By increasing the parameters $s$ and $R_0$ in $H_1$, we notice a significant enhancement of Hall conductivity peak, which originates from the Weyl nodes with opposite chirality, split from the nodal rings. Since we generally expect the higher order terms in $H_1$ to be small and vary from sample to sample, this explains the small anomalous Hall angle and a large variation of anomalous Hall conductivity (AHC) among the samples close to the charge neutrality point seen in experiments (see supplementary Note 4 and Supplementary Fig. 5 [57]). The shaded region in Fig. 5(f) illustrates the AHC range observed in experiments, which aligns with the theoretical calculations for the reasonable parameter sets.

**Summary**

We have uncovered the coupled evolution of the electrical and magnetic properties of Mn(Bi$_{1-x}$Sb$_x$)$_4$Te$_7$ through comprehensive studies on its single crystals using magnetization, magneto-transport, ARPES, and STM/STS measurements. Our investigation reveals that Sb doping at the Bi site not only tunes the system's chemical potential but also modulates the interlayer magnetic coupling. At $x \sim 0.27$, the chemical potential aligns closely with the charge neutrality point, where the charge carrier type transitions from electrons to holes, and the bulk carrier density reaches a minimum, approximately two orders of magnitude lower than that of pristine MnBi$_4$Te$_7$. The interlayer magnetic coupling evolves from AFM in MnBi$_4$Te$_7$ to FM for $0.15 \leq x \leq 0.20$ and $x \geq 0.45$ while reverting to AFM for $0.2 < x < 0.45$ before transitioning back to FM. The reemergence of AFM coupling at minimal carrier density highlights the critical role of the RKKY interaction in mediating interlayer FM coupling. For samples with a chemical potential near the charge neutrality point, we find that the AFM-to-FM transition induces an electronic structure transition, leading to the emergence of a type-II FM Weyl state. This Weyl state is evidenced not only by the chiral anomaly observed in magneto-transport measurements but also by Fermi surface reconstruction detected through the SdH oscillations. Furthermore, we find that this Weyl state also exhibits a 3D bulk QHE. Our theoretical studies reveal that this Weyl semimetal is highly anisotropic, with the Fermi velocity along the z-axis being an order of magnitude smaller than the in-plane velocity, which likely underpins the observed 3D quantum Hall effect (QHE). These findings position Mn(Bi$_{1-x}$Sb$_x$)$_4$Te$_7$ as a versatile platform for exploring diverse quantum states through Sb content control and uncover a unique type of FM Weyl semimetal state exhibiting a 3D QHE.

## III. METHODS

### A. Sample preparation and measurement

The melt growth method was employed to synthesize Mn(Bi$_{1-x}$Sb$_x$)$_4$Te$_7$ single crystals. High-purity Mn, Bi, Sb, and Te powders were thoroughly mixed in the stoichiometric ratio of Mn: Bi: Sb: Te = 1: (4-4x): 4x: 7, with $x$ ranging from 0 to 0.55, and loaded into carbon-coated quartz tubes, which were then sealed under high vacuum. The sealed quartz ampoules were placed in muffle furnaces and heated to 900°C for 5 hours to ensure melt homogeneity. Subsequently, the melt was rapidly cooled to an initial temperature $T_i$ within the range of 595°C to 605°C, with $T_i$ increasing alongside Sb concentration, as Sb-rich samples require higher growth temperatures. From $T_i$, the melt was slowly cooled at a rate of approximately 0.14°C per hour for three days, down to a final temperature $T_f$ within the 585°C to 595°C range. The slow cooling process was followed by annealing at $T_f$ for one day. Finally, the quartz ampoules were rapidly quenched in water to stabilize the metastable Mn(Bi$_{1-x}$Sb$_x$)$_4$Te$_7$ phase.

Following the synthesis process described above, high-quality single crystals of Mn(Bi$_{1-x}$Sb$_x$)$_4$Te$_7$ were obtained, and their phase purity was examined using X-ray diffraction (XRD) analysis. Powder XRD analysis of crushed crystals revealed that some samples contained minor intergrowths (< 5%) of (Bi,Sb)$_2$Te$_3$ or Mn(Bi,Sb)$_2$Te$_4$ phases. To ensure consistency and accuracy in our measurements, XRD scans were performed on each crystal used in the experimental investigations. Care was taken to confirm that impurity phases remained below 5%, minimizing their potential impact on the electronic and magnetic properties of the samples. Additionally, the compositions of the grown crystals were analyzed using Energy Dispersive X-ray Spectroscopy (EDS), which indicated that the actual Sb concentration was slightly lower than the nominal concentration by less than 3%. In this article, we use the nominal Sb concentration to label the samples. Supplementary Table S1 [57] in supplementary information summarizes the carrier type, density and mobility for most of the samples used in this study.

The magnetic properties of the samples were characterized using a Quantum Design Superconducting Quantum Interference Device (SQUID) magnetometer, capable of applying magnetic fields up to 7 T. Low-field magnetotransport measurements were performed over a temperature range of 2.5–300 K in a Quantum Design Physical Property Measurement System (PPMS), while extensive high-field magnetotransport measurements were conducted at the National High Magnetic Field Laboratory (NHMFL) in Tallahassee. All transport measurements were carried out using the standard four-probe method. For in-plane transport measurements, the samples were shaped into rectangles, with six leads affixed using silver epoxy to ensure reliable electrical contact. This configuration included one pair of current leads for applying in-plane current and two pairs of Hall voltage leads, designed to measure longitudinal resistivity ($\rho_{xx}$) and Hall resistivity ($\rho_{xy}$). The $\rho_{xx}$ and $\rho_{xy}$ data were extracted by symmetrizing and anti-symmetrizing the measurements taken under positive and negative magnetic fields, respectively. For out-of-plane (c-axis) transport measurements, the current leads were configured in a "C" shape on both cleaved sample surfaces to ensure a more homogeneous current distribution along the *c*-axis, with the voltage leads placed at the center of the "C" for accurate measurement.

ARPES measurements were carried out at BL-28A in Photon Factory (PF) with circularly polarized 90 eV photons using a micro beam spot of 12×10 μm$^2$ [73]. Samples were cleaved *in situ* in an ultrahigh vacuum of ~ 10$^{-10}$ Torr. Sample temperature during the ARPES measurements was kept at $T$ = 6 K. Energy and angular resolutions were set as 20 meV and 0.2°, respectively.

STM/STS measurements were carried out using a custom-made ultrahigh vacuum STM system [74]. The single crystals were cleaved at room temperature under UHV better than 5 × 10$^{−10}$ Torr and immediately transferred into the STM head kept at 4.5 K. STM/STS measurements were performed with PtIr tips at 4.5 K.

### B. Band structure calculations and theoretical methods

We performed the DFT calculations of electronic band structures for MnBi$_4$Te$_7$ in the FM phase. With the help of the Wannier function model, we also construct a low-energy Hamiltonian for three lowest energy conduction bands expanded around the A point in the Brillouin zone, which is labelled by $\vec{K}_A = \left(0,0,\frac{1}{2}\right)$ in the units of the reciprocal lattice vectors given by $b_1 = \frac{4\pi}{\sqrt{3}\,a_\parallel}\left(\frac{\sqrt{3}}{2}\hat{x} + \frac{1}{2}\hat{y}\right)$, $b_2 = \frac{4\pi}{\sqrt{3}\,a_\parallel}\hat{y}$, $b_3 = \frac{2\pi}{a_0}$ with $a_\parallel = 4.3615$ Å and $a_0 = 23.77$ Å. The three-band model is written as $H_{3b} = H_0 + H_1$, where $H_0$ is given by

$$H_0 = \begin{pmatrix} E_3 + a_3 k_\parallel^2 + a_{3z} k_z^2 & a_{23} k_- + i\, a_{23z} k_z k_+^2 & -i a_{13} k_z + i\, R_z k_z k_\parallel^2 \\ a_{23} k_+ - i\, a_{23z} k_z k_-^2 & E_2 + a_2 k_\parallel^2 + a_{2z} k_z^2 & -i\, a_{12} k_z k_+ \\ i\, a_{13} k_z - i\, R_z k_z k_\parallel^2 & i\, a_{12} k_z k_- & E_1 + a_1 k_\parallel^2 + a_{1z} k_z^2 \end{pmatrix}$$

and $H_1$ is written as

$$H_1 = \begin{pmatrix} 0 & 0 & R_1(k_x^3 - 3 k_x k_y^2) - i\, R_2(3 k_x^2 k_y - k_y^3) \\ 0 & 0 & s\, k_-^2 \\ R_1(k_x^3 - 3 k_x k_y^2) + i\, R_2(3 k_x^2 k_y - k_y^3) & s\, k_+^2 & 0 \end{pmatrix}$$

where $k_\parallel = (k_x^2 + k_y^2)^{1/2}$ and $k_z$ label the momentum expanded from the A point, namely $\mathbf{k} = \vec{K}_A + \delta\mathbf{k}$ with $\delta\mathbf{k} = \sum_{i=x,y,z} k_i \hat{e}_i$ ($\hat{e}_i$ is the unit-vector along the $i$th direction), and $k_\pm = k_x \pm i\, k_y$. The parameters are given by $E_1 = 0.06445$ eV, $E_2 = 0.07718$ eV, $E_3 = 0.08638$ eV, $a_1 = 82.1027$ eV Å$^2$, $a_2 = 33.6563$ eV Å$^2$, $a_3 = 7.32603$ eV Å$^2$, $a_{12} = 0.635236$ eV Å$^2$, $a_{13} = 0$, $a_{23} = 0.0207114$ eV Å, $a_{23z} = 51.7824$ eV Å$^3$, $R_1 = 15.769$ eV Å$^3$, $R_2 = 5.88178$ eV Å$^3$, $s = 1.38879$ eV Å$^2$, $R_z = 49.7405$ eV Å$^3$, $a_{1z} = 14.6301$ eV Å$^2$, $a_{2z} = 23.2978$ eV Å$^2$, $a_{3z} = 6.14636$ eV Å$^2$.

**Acknowledgments:**

**Funding**: This work was primarily supported by the US Department of Energy under grant DE-SC0019068 (support for personnel, material synthesis, magnetic measurements, magnetotransport measurements and data analyses). Z.Q.M. also acknowledges support from JSPS Invitational Fellowship for Researcher in Japan and AIMR GI$^3$ Lab for facilitating the collaborative experiment (ARPES & STM/STS) conducted at Tohoku University. Y.C. acknowledges the support from JSPS KAKENHI Basic Science A (22H00278). S.H.L. acknowledges the support from the Pennsylvania State University Two-Dimensional Crystal

Consortium–Materials Innovation Platform (2DCC-MIP), which is supported by NSF Cooperative Agreement No. DMR-2039351. J.W.S. acknowledges the support from DOE grant (DE-SC0014208) for the DFT calculations of the density of state across the magnetic transition. T.S. acknowledges the support from JST-CREST (No. JPMJCR18T1). A.H. thanks GP-Spin and JSPS for financial support. A portion of this work was performed at the National High Magnetic Field Laboratory, which is supported by National Science Foundation Cooperative Agreement No. DMR-2128556 and the State of Florida